\documentclass[12pt,aps,prb,preprint,superscriptaddress]{revtex4}

\usepackage{amsmath}    
\usepackage{graphicx}   

\begin{document}

\title{Public exhibit for demonstrating the quantum of electrical conductance}
\author{E. H. Huisman}
\altaffiliation[Current address: ]{Columbia University, Center for Electron Transport in Molecular Nanostructures, 500 W 120th St., New York, NY-10027, New York, United States of America}  
\email{ehh2125@columbia.edu}   
\author{F. L. Bakker}
\affiliation{University of Groningen, Zernike Institute for Advanced Materials, Nijenborgh 4, NL-9747 AG, Groningen, The Netherlands }
\author{J.P. van der Pal}
\affiliation{University of Groningen, Zernike Institute for Advanced Materials, Nijenborgh 4, NL-9747 AG, Groningen, The Netherlands }
\author{R. M. de Jonge}
\affiliation{University of Groningen, Science Linx, Nijenborgh 9, NL-9747 AG, Groningen, The Netherlands }
\author{C. H. van der Wal}
\affiliation{University of Groningen, Zernike Institute for Advanced Materials, Nijenborgh 4, NL-9747 AG, Groningen, The Netherlands }

\date{\today}

\begin{abstract}
We present a new robust setup that explains and demonstrates the quantum of electrical conductance for a general audience and which is continuously available in a public space. The setup allows users to manually thin a gold wire of several atoms in diameter while monitoring its conductance in real time. During the experiment, a characteristic step-like conductance decrease due to rearrangements of atoms in the cross-section of the wire is observed. Just before the wire breaks, a contact consisting of a single atom with a characteristic conductance close to the quantum of conductance $G_0=\frac{2e^2}{h}$ can be maintained up to several seconds. The setup is operated full-time, needs practically no maintenance and is used on different educational levels.
\end{abstract}

\maketitle

\section{Introduction}
Reducing the size of electronic components has been a main ingredient for innovation in the electronics industry over the last half century. Interestingly, by reducing the size of electronic structures, the physics needed to describe the electron's motion changes. A characteristic length scale marking the border between two different physical regimes is the average distance over which an electron travels before it gets scattered, i.e. the electron's mean-free path $L_e$. For example in copper $L_e$ = 39 nm (1 nm = 10$^{-9}$ m) at room temperature.\cite{kittel} In fact, this length is comparable to the size of today's smallest electronic components.\cite{intel}   Hence, one might wonder what happens to the electrical conductance of a conductor when reducing its size below $L_e$? This simple yet intriguing question was answered in 1988 when improvements in semiconductor technology allowed the fabrication of nanoscale devices with very precisely controllable dimensions.\cite{wees,pepper} When $L<<L_e$, one enters the so-called ballistic regime. In this regime electrons still bounce off the walls of the conductor, but hardly scatter between themselves.  The 1988 experiments showed that for a ballistic piece of conductor, although hardly any electron-electron scattering is expected, still a finite electrical conductance $G=\frac{1}{R}=\frac{I}{V}$ is measured. Furthermore, the conductance decreased in a step-like fashion while continuously decreasing the conductor's cross-section. This differs from macroscopic conductors, whose conductance scales continuously with its cross-sectional area (Ohm's law). Most intriguingly, the size $G_0$ of each conductance step turns out to be material independent. It is an exact universal constant, or quantum, constituted purely out of fundamental constants,
and it is called the quantum of conductance $G_0=\frac{2 e^2}{ h}= \: \frac{1}{12.9\:k\Omega}$, where $e$ is the electron's charge and $h$ is Planck's constant.

The sub-area of condensed matter physics studying the physics of electron transport at these length scales is called mesoscopic physics. Concepts developed in this field during the 1980s have greatly shaped our current understanding of electron transport in small electronic devices. Although well-described nanometer scale physics has entered our daily-life, a general audience mostly lacks basic knowledge of phenomena at this scale. We believe there is a need for educational tools to bridge this gap. In this paper, we present a public demonstration setup in which users explore one of the central concepts of the field of mesoscopic physics, the quantum of conductance. This exhibit thereby illustrates that the physics of quantum and nano systems differs qualitatively from the physics that governs our macroscopic world. As a consequence, it is suited for provoking fascination and enthusiasm for nanoscale physics.

\section{The Physics of Quantized Conductance}

The physics that underlies conductance quantization results from quantum confinement of electrons. To see how quantum confinement applies to a nano-scale conductor, let us consider a piece of 2-dimensional conductor of width $W$ and length $L$  connected to two pieces of bulk conductor (see Figure 1a).  When $L<<L_e$, one enters the so-called ballistic regime. In this regime electrons bounce on the walls of the conductor, but hardly scatter between themselves and can thus be regarded as freely moving particles. In the \textit{transverse} direction, the electrons interfere between the walls of the constriction and form discrete quantum states or modes at energies $E_{nx}$. Here, $n={1,2,3,4,...}$ is the mode index for each wave harmonic (depicted in Figure 1b). In other  words, standing waves between the walls of the channel are formed similar to the particle-in-a-box model, demonstrating the wave-like nature of electrons. However, the conductance is determined by electrons that propagate \textit{along} the constriction. In this direction, within each state with index $n$, the electrons are still free to propagate. The 1988 experiments show that while continuously decreasing $W$ of the constriction, the conductance decreases in a step-like fashion.  This can be understood by realizing that the only discrete states that are actually occupied by electrons are those that lie below the energy (Fermi energy) of the conduction electrons in the bulk conductors. These conduction electrons in the metal are characterized by a wavelength called the Fermi wavelength $\lambda_F$. Therefore, the index $n$ of the highest occupied electron mode is given by $N_{max} \approx \frac{2W}{\lambda_F}$. Hence, the number of occupied modes can be controlled by changing the width of the channel. Each time the channel width is reduced by $\frac{\lambda_F}{2}$, $N_{max}$ is reduced by one mode. Thus, as the channel gets narrower, the number of occupied levels drops in a step-like fashion.

Strikingly, each of the $N_{max}$ modes in the constriction contributes exactly one quantum of conductance to the channel's total conductance. To probe the junction's conductance $G=\frac{I}{V}$ an energy offset between electrons in the two bulk conductors at each side of the constriction is created by applying a small bias voltage $V$ while measuring the resulting current $I$.  The current carried by each mode can be determined by counting the number of particles carrying a charge $e$ over the energy interval given by the applied bias $eV$:\cite{vanhouten} $I_n=2e\int_0^{eV}\rho_n(E)  v_n(E)  dE$.  Here, $\rho_n(E)$ is the 1-D density of states for electrons moving in mode $n$ in the forward direction, and $v_n(E)$ is the group velocity of an electron in mode $n$ at energy $E$. The 2 is due to the spin degeneracy: each mode can carry an electron with spin up and spin down.  Strikingly, in the 1-dimensional case, the product of $\rho_N(E) v_N(E)$ turns out to be exactly equal to the inverse of Planck's constant, $1/h$ (see Ref.~\onlinecite{vanhouten}).  Therefore, $I_n=(2e^2/h)V$  and the conductance  of each mode is a constant $G_0=2 e^2/h$ independent of $E$, $V$ or mode index $n$. Hence, by continuously decreasing the width $W$ of the conductor, a step-like decrease in conductance is observed at equidistant intervals $G_0$, see Figure 1c.
For a full derivation of the quantum of conductance we refer to references 6 and 7. 

The initial observation of conductance quantization was made at temperatures of about 1 Kelvin with channels made out of hi-tech materials (epitaxial layers of semiconductors, grown in ultra-high vacuum). These materials feature a relatively long Fermi wavelength $\lambda_F$ = 42 nm. To create the constriction, an advanced patterning technique called electron-beam lithography was used. These conditions are elaborate and  expensive and not suitable for a permanent public exhibit. However, in the 1990s it was found that atomically-sized metal wires with a diameter of a few atoms often have conductances near integer values of the quantum of conductance.\cite{agrait} The archetypical example is gold. By applying a stress in the longitudinal direction of the golden wire, its atoms rearrange to form a longer and thinner wire. This thinning can be continued up to the point where the thinnest part spans only one atom.  Although steps occur at virtually all conductance values, preferred conductances appear at integers of $G_0$. Most notably, just before the wire breaks, a plateau very close to 1 $G_0$ is observed.  Conductance measurements of gold atomic contacts in a high resolution transmission electron microscope confirmed that wires with a conductance of 2 $G_0$ and 1 $G_0$ correspond to a configuration having two and one atom in the smallest diameter of the wire. \cite{takayanagi}  The diameter of a gold atom in a lattice (2.5 \AA) is about half the value of the Fermi wavelength ($\lambda_F =0.52 nm$).  Therefore, a single atom indeed forms a constriction as described in Figure 1 with $N_{max}=1$.  However, it is less clear how atomic contacts make a constriction with $N_{max}>1$.  When thinning an atomic gold wire, the diameter of the wire decreases in a step-wise fashion, atom by atom. Therefore, the preferred conductances at multiple integers of the quantum of conductance can also be explained as arising from the discrete number of atoms in the wire (each contributing 1 $G_0$) instead of transversal quantum confinement in a constriction of varying width. For gold, it turns out both explanations are correct, as discussed below.

Studies on atomically-sized contacts made from different metals showed interesting nuances. For example, not all metals form single atom contacts with a conductance close to 1 $G_0$. In contrast to gold, Transition metals often show a plateau before breaking around  non-integer values $>> 1 \: G_0$. \cite{agrait} This difference seems to be correlated with both the shape and occupancy of the outer valence orbitals of the atom. A gold atom has a spherically symmetrical outer valence orbital (6s) with only one electron in it. This orbital dominates charge transport near the Fermi energy. In contrast, a transition metal can have multiple electrons in strongly directional outer orbitals. Also, not all metals form single atom contacts with a conductance near 1 $G_0$ carried by a \textit{single} quantum mode. This can be understood by realizing that a  mode as depicted in Figure 1 does not have to be fully transparent and might contribute only a fraction of the quantum of conductance in systems where back-reflection of electron waves in the channel can occur. \cite{NOTE1} For example, for aluminum, a contact with a conductance value close to 0.8 $G_0$ appears frequently just before breaking. However, it was found that this conductance is actually carried by three partly transmissive modes. \cite{agrait} For gold, electrical (shot) noise measurements did confirm that the current in a single atom contact is indeed carried by an almost fully transparent single quantum mode. \cite{brom}  Therefore, the 1 $G_0$ plateau is an elegant demonstration of conductance quantization; a gold atom acts as a single fully transparent waveguide for electron waves. It directly demonstrates what conductance value occurs in ultimately thin and clean nanowires.

\section{Description of the Public Exhibit}

Simple experimental realizations of the gold wire experiments have been proposed before. One can even observe the quantum of conductance by disrupting two gold wires attached to a table by tapping the table top \cite{costa, tuominen} or by opening an electromagnetic relay.\cite{dublin} However, these experiments still need to be supervised in order to be performed by a lay person in a short time span. This supervision is a rather elaborate and expensive task and greatly limits the accessibility of these experiments to a general audience. Furthermore, the time span over which the quantization is observed is on the order of milliseconds or less. This requires detection with a fast oscilloscope and does not allow an observation at the time scale of a human observer. In this paper, we report a robust, low maintenance setup that allows a user to perform the gold wire experiment in real time. The setup is permanently accessible in a public space where a user can operate it without supervision. The budget available for equipment to realize the setup was about \$13,000. Below, we give a description of the technique used to make atomic gold contacts and a description of the design of the setup.

In order to create atomic contacts, we decided to use the lithographically defined mechanical break junction technique. Figure 2a shows a schematic drawing of this technique. A gold wire is defined on top of a flexible, insulating substrate and is designed to be notched in the center to create a weak point. The substrate is clamped in a three-point bending configuration.   The sample bends by moving the central pushing rod up while holding the outer ends of the sample using counter supports. This results in a shear force on the wire, thereby elongating the wire and reducing its diameter. Mechanical break junctions feature a short distance (about 2 $\mu$m) over which the gold wire is suspended.  Due to its geometry, the ratio between the pushing rods displacement $\Delta Z$ and inter electrode displacement $\Delta d$, $r = \Delta d/\Delta Z$ is much smaller than one. \cite{vrouwe} For our junctions, we estimate $r \approx 2 \cdot 10^{-4}$. Due to this small translational attenuation factor, the small `mechanical loop' and high symmetry, mechanical break junctions are almost insensitive to vibrations. Therefore, the lifetime of the single atom contact is usually not limited by vibrations and allows the user to observe the formation of a single atom contact while slowly breaking a thin gold wire in about one minute.

Initially, we fabricated junctions using phosphor bronze substrates and electron beam lithography, as in recent research projects with these systems. \cite{trouwborst} However, in order to fabricate samples cheaper and more efficiently, we developed a simpler fabrication method on Cirlex\textregistered $\:$ substrates \cite{KATCO} using optical lift-off based lithography. We use a deep-UV mask aligner and vacuum evaporation to pattern  gold break junctions of 120 nm thickness. Four junctions are simultaneously defined on top of a polished Cirlex\textregistered $\:$ substrate of 22~mm x 20~mm x 0.51~mm (length x width x thickness). The patterns include contact pads for the spring-loaded pins depicted in Figure 2a. In order to get a suspended central constriction of the junctions, the samples are exposed to an oxygen plasma that etches away a small amount of substrate material. Finally, the four junctions are separated using regular scissors, see Figure 1c. Figure 2d displays a scanning electron micrograph of a finished junction.  We experienced that a junction needs to be replaced after 3-6 months at full-time operation.  A large number of samples can be made by spending less than a day in any facility equipped for microelectronic device fabrication. Thus, an exhibit at a site without its own device fabrication facilities can be supplied with a multi-year stock of samples (say 20 samples) for less than \$1000.

Figure 3 shows a photograph and close-ups of the setup as realized in the entrance hall of our Science Faculty building. The  setup is controlled via a touch screen and a turning handle.  The break junction, spring contacts, bending bench and the mechanical transmission are displayed in a transparent case. Thus, while operating the setup, the motion of mechanical parts and in particular the bending of the break junction are clearly visible to the user. In the experiment, the user directly controls the pushing rod using a turning handle. The handle's rotation is translated in a linear motion of the pushing rod via a precision micrometer (50 $\mu$m translation per rotation). A low gear-ratio worm drive (1:20) between the micrometer and the turning handle further attenuates the relation between user motion and inter electrode displacement. As a consequence, $\Delta d$ can be controlled with an impressive precision: A full turn of the handle results in a 2.5 $\mu m$ translation of the pushing rod. This translation results in an electrode-electrode displacement $\Delta d$ of roughly 0.5 nm, the equivalent of 2 gold atoms, while the user can easily control a small fraction of a full turn.

The conductance of a break junction is determined by applying a constant bias voltage $V$ and measuring the current $I$ through the junction using a home-built trans-impedance amplifier (Figure 2b). A central component in the setup is a regular personal computer equipped with a National Instruments data acquisition board (M6221). The acquisition board applies the bias voltage $V$ and records the output of the current meter at 100.000 samples per second. Furthermore, it actively protects the junction and the rest of the setup against damage that might occur if a user would rotate the handle outside the designed operation range. To this end, it reads out a motion (Hall) sensor installed on the handle, which detects the direction of handle rotation. We use this signal together with the measured conductance values as inputs to a feedback mechanism which controls a coaxial magnetic coupler between the handle and the worm drive.  If the user is turning in the undesired  direction (closing when opening is required and vice versa) for a considerable time, the computer will decouple the handle from the central axis. Also, two switches are installed just before an upper and lower limit for the micrometer. When a switch is hit by the micrometer, the handle is automatically decoupled.

A software interface consisting of a Microsoft PowerPoint presentation embedded in a National Instruments Labview code guides the user through the experiment. The interface has three parts. During the first part, a slide show introduces the context of nano-electronics and the relevant physical concepts. In the second part the actual experiment is performed, and the computer presents a graph with conductance versus time (as displayed in Figure 3a) while the user turns the handle.  After breaking, the user needs to close the junction.  By closing the junction, the two electrodes `self-repair' into a single gold wire \cite{trouwborst} and  the user can choose whether he/she wants to redo the experiment. For a single junction,  this opening and closing cycle can be repeated $>$ 1000 times. When proceeding to the third part, a slide show further explains the observations. Several parts of this setup (mechanical, electronics and software) were realized by undergraduate students in projects that were part of their curriculum, with graduate students having a primary role in the supervision and project management.

\section{Results and Discussion}

Figure 4a shows five typical curves obtained in an experiment by rotating the handle at a constant speed of approximately 1 turn per second ($\sim 0.5$ nm per second). The conductance $G$ of the wire is expressed in units of the quantum of conductance ($G_0=2e^2/h=77.5 \: \mu S$). For each trace approximately 10 turns ($\sim$ 5 nm) are needed to break the junction. Clearly, while the junction is stressed in a continuous fashion, the conductance drops in a step-like fashion due to rearrangements of the atoms in the junction. Plateaus occur at virtually any conductance value $> 1 \: G_0$. However, plateaus appear more frequently at a conductance near 1 $G_0$,  indicating that a relatively stable configuration, i.e. a single atom contact, is formed. The breaking of the single atom contact is observed as an abrupt jump in the conductance at the end of the 1 $G_0$ plateau. If one now starts to close the junction again, the wires ends only make contact after going back a few turns, thus revealing a hysteresis in the dependence of conductance on $\Delta Z$. \cite{trouwborst} As soon as the wire ends touch each other again, we typically observe a very rapid increase in the conductance to a value of several $G_0$ (data not shown).

To illustrate the stability of the setup, we measured the lifetime of 108 single atom contacts while slowly rotating the pushing rod ($\approx$ 2 rpm) and stopping further turning as soon as a 1 $G_0$ plateau was observed. Figure 4b shows a lifetime histogram of all contacts. The lifetime of the contact varies over 3-4 orders of magnitude centered around 1~s. Previously it was shown that using similar junctions of a similar geometry at low temperatures, single atom contacts can be maintained stable for hours.\cite{agrait,trouwborst} This indicates that the intrinsic mobility of the gold atoms at room temperature limits the lifetime of a single atom contact using the mechanical break junction technique.

Our setup has been part of a permanent exhibition in the entrance hall of the faculty building of Natural Sciences of our university since september 2008. The exhibition aims to display current research of the faculty to a broad audience.  Within this exhibition, our setup is used as an educational tool on two levels. The setup is used as a stand-alone exhibit, where students, faculty staff and visitors can measure the conductance of a single gold atom in a few minutes without any supervision. The user will perform the break junction experiment and most importantly, get familiarized with the quantum of conductance. Second, the setup is used as a guideline to exemplify research in nanoscience. Small groups (2-5 persons) of high school or bachelor students and a supervisor (often a graduate student) perform the experiment and go through the steps of sample preparation in the cleanroom to illustrate nano-fabrication. Sample inspection using a scanning electron microscope is used to visualize a break junction. By zooming in on the sample using this microscope one can illustrate the `powers of ten' to be overcome if one wishes to engineer at the nanometer scale.

Future plans for the setup include implementing an internet interface to the setup. Users can then send the results of an experiment to a personal e-mail address. This allows adding a third layer of usage in which scholars do the experiment while visiting the exhibit and can work-out and analyze the results on a computer, either during the visit to our faculty, or later at home or at school. Analysis and workout can be supported by on-line material on the website of the exhibit.\cite{website} Furthermore, we plan to monitor the results of the setup over longer periods of time, and to collect statistics of the conductance values that occur. This allows, for example, to investigate the long-term behavior and aging effects of these junctions with broad audience participation and real time reporting on the website. For future setups, we estimate that fabrication costs can be reduced to approximately \$7,000.

In summary, we presented a robust setup that allows a general audience to perform an experiment demonstrating the quantum of conductance. The setup is permanently available  in the entrance hall of our Science Faculty building for outreach purposes.

\begin{acknowledgments}
The authors like to acknowledge Maarten Inklaar and Gerja Evers for their help in designing the slide show, Harry van Driel, Bert van Dammen, Koos Duijm, Wigger Jonker, Bernard Wolfs and Marc Petstra for their help in realizing the setup, Hector Vazquez for usefull comments and Bart van Laar for Science Linx funding.
\end{acknowledgments}

\begin{figure}[ht]
	\centering
		\includegraphics[width=0.6\textwidth]{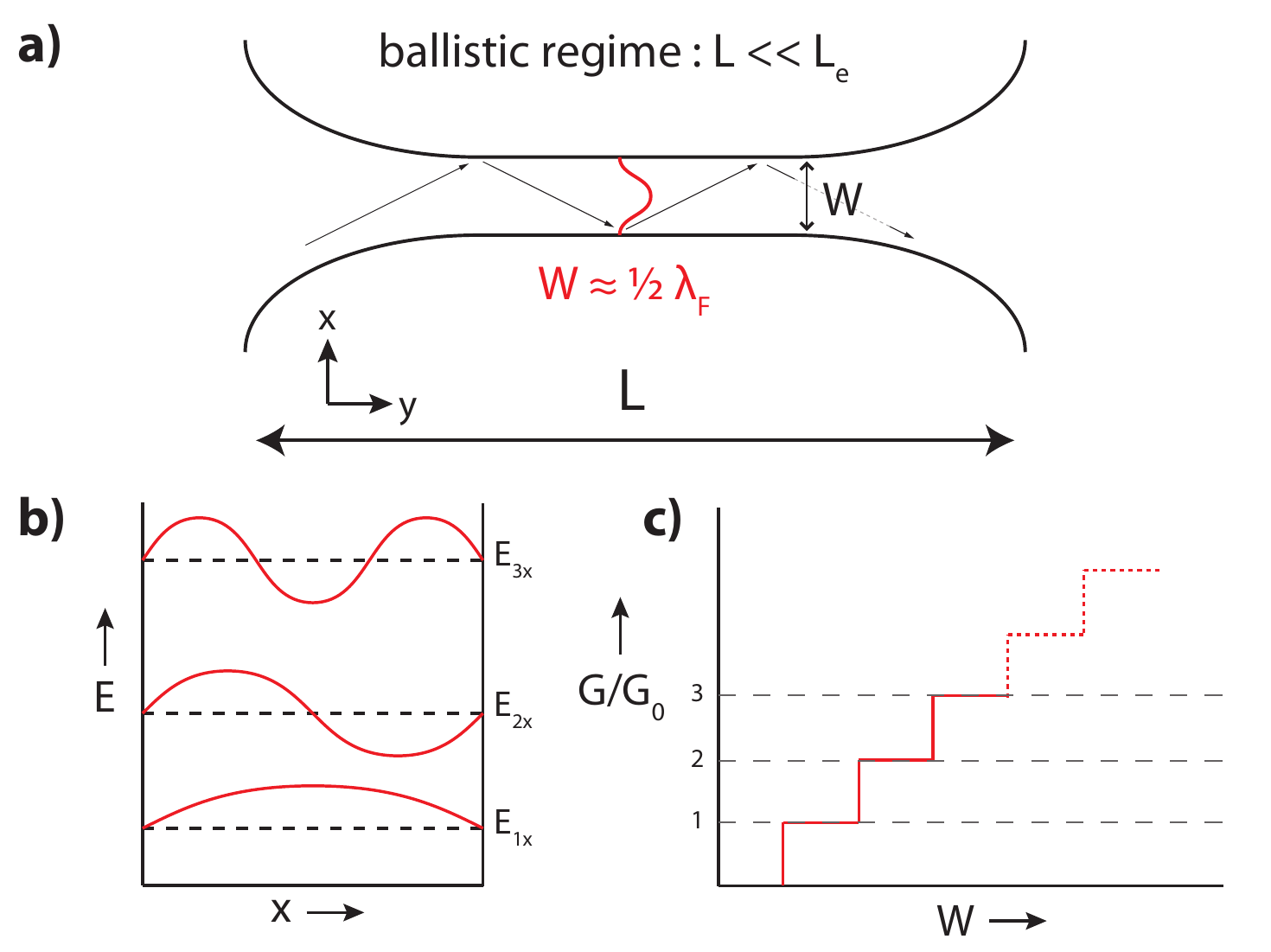}
	\caption{\textbf{a}~Schematic representation of a 2-dimensional ballistic piece of conductor of length $L$ and width $W$. In the ballistic regime, $L$ is much smaller than the mean free path of the electrons $L_e$. The arrows depict a classical presentation of a trajectory of an incoming electron. However, at these length scales electron propagation must be described as quantum waves. These quantum waves interfere between the walls of the constriction, and standing wave patterns develop in the \textit{transverse} direction $x$. The confinement thus gives rise to a discrete set of energy modes $E_{nx}$, depicted in \textbf{b}. Here, $n$ is the mode index and indicates the number of 'half' electron waves (1/2 times the Fermi wavelength, $\lambda_F$) that can fit in $W$. Within each transversally confined level, the electrons propagate freely \textit{along} the direction of the wire. Hence, the propagating electrons behave as particle moving in 1-dimension only.  \textbf{c}~Strikingly, each mode $n$ contributes exactly one quantum of conductance to the total conductance of the wire. When continuously reducing $W$ of the constriction, the conductance will decrease in a step-like fashion. Each time $W$ is reduced by $\frac{\lambda_F}{2}$, the conductance drops with one quantum of conductance. This phenomenon is called  conductance quantization. }
\end{figure}

\begin{figure}[ht]
	\centering
		\includegraphics[width=0.6\textwidth]{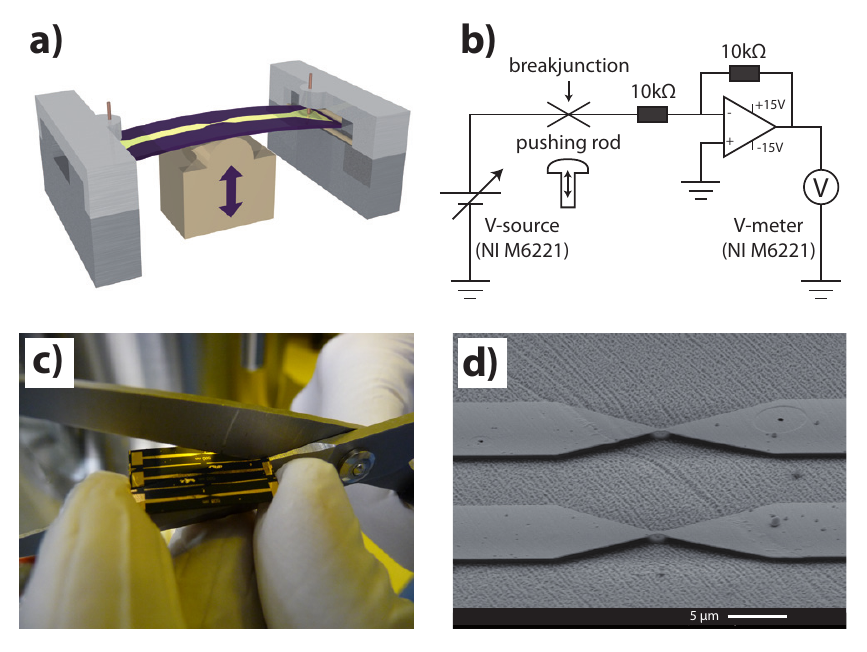}
	\caption{\textbf{a}~Schematic representation showing the mechanical break junction clamped in a three-point bending configuration. The pushing rod's motion results in a lateral stress on the lithographically patterned gold wire, thereby thinning and eventually breaking it. \textbf{b}~Simplified electronic scheme of the entire setup. The voltage (V-source) is applied  by a computer via a data acquisition card (NI M6221). This card also reads the output of the current meter (realized as a V(olt)-meter which probes the output of a trans-impedance amplifier). The pushing rod is manually controlled using a turning handle. The computer is connected to a touch screen.  \textbf{c}~Four junctions are simultaneously fabricated on a Cirlex(r) substrate using lift-off based photo lithography. Individual samples can be isolated using scissors. \textbf{d}~Scanning electron micrograph showing two suspended mechanical break junctions after device fabrication.}
\end{figure}

\begin{figure}[ht]
	\centering
		\includegraphics[width=0.6\textwidth]{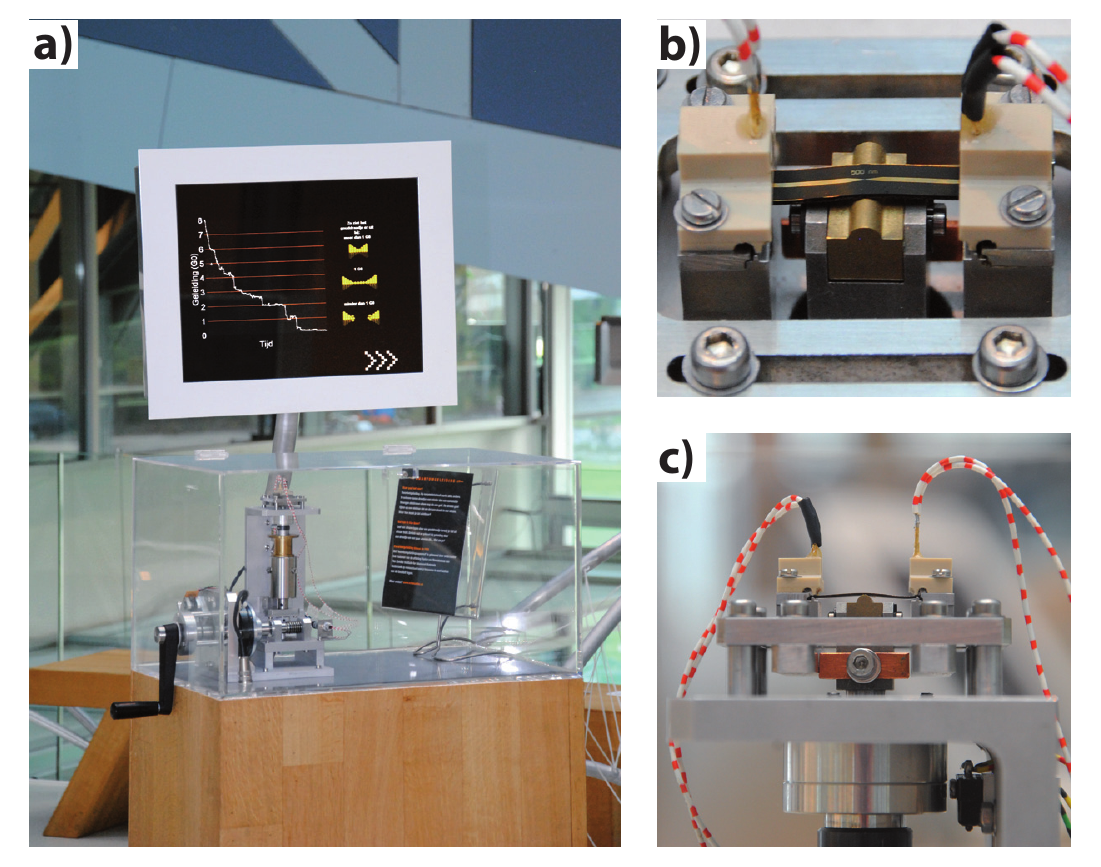}
	\caption{ \textbf{a}~Photograph of the entire setup as realized in the entrance hall of our faculty building. The user can operate the setup using a turning handle and a touch screen. \textbf{b,c}~Close-up of the mechanical part of the setup showing a top and side view  of the three-point bending bench with a break junction. The distance between the wire connections on the left and right side is about 2.5~cm.}
\end{figure}

\begin{figure}[ht]
	\centering
		\includegraphics[width=0.6\textwidth]{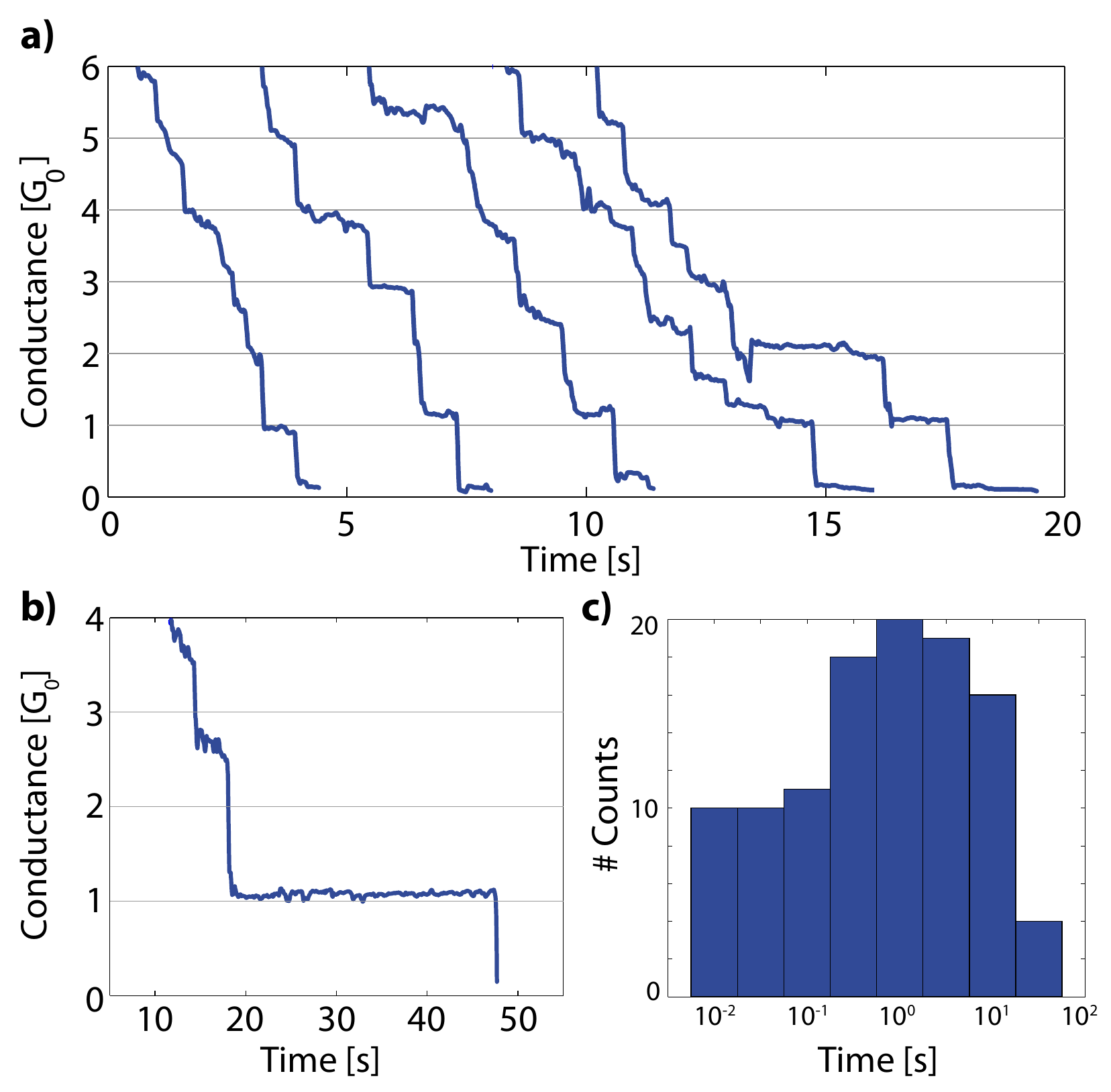}
	\caption{\textbf{a}~Five typical conductance versus time traces when turning the handle at 1 turn per second. The applied voltage is 100 mV. 
\textbf{b,c}~The stability of the mechanically controllable break junction technique is illustrated when turning the handle at low speeds ($\sim$ 2~rpm) and holding the handle once a conductance of 1 $G_0$ is reached. In \textbf{b}, a trace featuring a relatively long 1 $G_0$ plateau is plotted. When repeating the experiment 108 times, the single atom contact holding time (= time the conductance remains between 1.5 and 0.5 $G_0$) features a wide distribution centered around 1 second. At room temperature, the lifetime of a single atom contact is limited by the diffusion of gold atoms. }
\end{figure}


\begin{thebibliography}{5}

\bibitem{kittel} C. Kittel, ``Introduction to solid state physics" 8th edition Wiley

\bibitem{intel} J. Hicks, D. Bergstrom, M. Hattendorf, J. Jopling, J. Maiz, S. Pae, C. Prasad and J. Wiedemer,  ``45nm Transistor Reliability", Intel Technology Journal {\bf 12}, 131-144 (2008).

\bibitem{wees} B.J. van Wees, H. van Houten, C.W.J. Beenakker, J.G. Williamson, L.P. Kouwenhoven, D. van der Marel and C.T. Foxon, ``Quantized conductance of point contacts in a two-dimensional electron gas," Phys. Rev. Lett. {\bf 60}, 848-850 (1988).

\bibitem{pepper} D.A. Wharam, T.J. Thornton, R. Newbury, M. Pepper, H. Ahmed, J.E.F. Frost, D.G. Hasko, D.C. Peacock, D.A. Ritchie and G.A.C. Jones, ``One-dimensional transport and the quantisation of the ballistic resistance," J. Phys. C {\bf 21}, L209-L214 (1988).

\bibitem{vanhouten} H. van Houten and C.W.J. Beenakker, ``Quantum Point Contacts," Physics Today {\bf 49},22-27 (1996).

\bibitem{Datta} S. Datta, ``Electronic transport in mesoscopic systems," Cambridge University Press (1995).

\bibitem{harmans} C. Harmans, ``Mesoscopic Physics, an introduction," TUDelft (1997).

\bibitem{agrait} N. Agra\"{\i}t, A.L. Yeyati and J.M. van Ruitenbeek, ``Quantum properties of atomic-sized conductors," Physics Reports {\bf 377}, 81-279 (2003).

\bibitem{takayanagi} H. Ohnishi, Y. Kondo and K. Takayanagi ``Quantized conductance through individual rows of suspended gold atoms" Nature {\bf 395}, 780-783 (1998).

\bibitem{NOTE1} Modes contributing a fraction of the quantum of conductance arise when an incoming electron is reflected inside the ballistic channel, e.g. due to interacting with lattice vibrations or scattering off impurities.

\bibitem{brom} H.E. van den Brom and J.M. van Ruitenbeek, ``Quantum Suppression of Shot Noise in Atom-Size Metallic Contacts," Phys. Rev. Lett. {\bf 82} 1526-1529 (1999).

\bibitem{costa} J.L. Costa-Kr\"amer, N. Garcia, P. Garcia-Mochales and P.A. Serena, ``Nanowire formation in macroscopic metallic contacts: quantum mechanical conductance tapping a table top," Surf. Science {\bf 342}, L1144-L1149 (1995).

\bibitem{tuominen} E.L. Foley, D. Candela, K.M. Martini and M.T. Tuominen, ``An undergraduate laboratory experiment on quantized conductance in nanocontacts," Am. J. Phys. {\bf 67}, 389-393 (1999).

\bibitem{dublin} F. Ott and J. Lunney, ``Quantum Conduction: a Step-by-Step Guide," Europhysics News {\bf  January/February}, 13-16 (1998).

\bibitem{vrouwe} S.A.G. Vrouwe, E. van der Giessen, S.J. van der Molen, D. Dulic, M.L. Trouwborst and B.J. van Wees, ``Mechanics of lithographically defined break junctions," Phys. Rev. B {\bf 71} 0353131-0353137 (2005).

\bibitem{trouwborst} M.L. Trouwborst, E.H.Huisman, F.L. Bakker, S.J. van der Molen and B.J. van Wees, ``Single Atom Adhesion in Optimized Gold Nanojunctions," Phys. Rev. Lett. {\bf 100} 1755021- 1755024  (2008).

\bibitem{KATCO} A sheet of Cirlex\textregistered $\:$ was purchased from Katco Ltd, see http://www.katco.uk.com

\bibitem{website} http://www.rug.nl/sciencelinx

\end{thebibliography}
\end{document}